# Model Learning: Primal Dual Networks for Fast MR imaging

Jing Cheng[1], Haifeng Wang[1], Leslie Ying[3], Dong Liang[1, 2(✉)]

[1] Paul C. Lauterbur Research Center for Biomedical Imaging, Shenzhen Institutes of Advanced Technology, Chinese Academy of Sciences, Shenzhen, Guangdong, China
[2] Research center for Medical AI, Shenzhen Institutes of Advanced Technology, Chinese Academy of Sciences, Shenzhen, Guangdong, China
Dong.Liang@siat.ac.cn
[3] Departments of Biomedical Engineering and Electrical Engineering, University at Buffalo, the State University of New York, Buffalo, NY 14260 USA

**Abstract.** Magnetic resonance imaging (MRI) is known to be a slow imaging modality and undersampling in k-space has been used to increase the imaging speed. However, image reconstruction from undersampled k-space data is an ill-posed inverse problem. Iterative algorithms based on compressed sensing have been used to address the issue. In this work, we unroll the iterations of the primal-dual hybrid gradient algorithm to a learnable deep network architecture, and gradually relax the constraints to reconstruct MR images from highly undersampled k-space data. The proposed method combines the theoretical convergence guarantee of optimization methods with the powerful learning capability of deep networks. As the constraints are gradually relaxed, the reconstruction model is finally learned from the training data by updating in k-space and image domain alternatively. Experiments on in vivo MR data demonstrate that the proposed method achieves superior MR reconstructions from highly undersampled k-space data over other state-of-the-art image reconstruction methods.

**Keywords:** MR reconstruction, Primal dual, Deep learning.

## 1 Introduction

Accelerating magnetic resonance imaging (MRI) has been an ongoing research topic since its invention in the 1970s. Among a variety of acceleration techniques, compressed sensing (CS) has become an important strategy during the past decades [1]. In general, the imaging model of CS-based methods can be written as

$$\min_{m} \frac{1}{2} \|Am - f\|_2^2 + \lambda \|\Psi m\|_1 \quad (1)$$

where the first term is the data consistency and the second term is the sparse prior. $\Psi$ is a sparse transform, such as wavelet transform or total variation, $m$ is the image to be reconstructed, $A$ is the encoding matrix, $f$ denotes the acquired k-space data.

Although CS-based methods can achieve high performance with many theoretical guarantees, it is challenging to determine the numerical uncertainties in the model such as the optimal sparse transformations, sparse regularizer in the transform domain, regularization parameters and the parameters of the optimization algorithm.

Recently, deep learning has demonstrated tremendous success and has become a growing trend in the general field of data analysis [2]. It also has been introduced in MR reconstruction and shown potential to significantly speed up MR acquisition and improve image quality [3-11]. Deep learning-based MR reconstruction can be generally divided into two categories: data-driven [3-7] and model-driven [8-11]. Data-driven methods directly learn an end-to-end mapping between the input and the output with little prior knowledge through a predesigned network architecture. However, they usually require a large size of training data and long training time. Model-driven methods unroll the iterations of an optimization algorithm to a deep network to learn the constraints and parameters in the reconstruction model from training data. As a result, such networks can perform well with a smaller size of training data.

In this work, we start from the traditional CS-MRI reconstruction and aim to maximize the potential of deep learning and model-based reconstruction. Using the primal dual framework as an example, we explain how to unroll the iterations of a reconstruction process to a learnable deep network architecture. The prior of the to-be-reconstructed image is obtained by the trained networks and the data consistency is also maintained through updating in k-space for the reconstruction, which is not typical for most existing deep learning MR reconstruction methods. Our work can be considered as a preliminary study on connecting the model-driven methods with data-driven methods.

## 2    Primal Dual Networks

### 2.1    PDHG-CSnet: learning operator and parameters

The primal dual hybrid gradient algorithm, also known as Chambolle-Pock (CP) algorithm [12], has been applied on several imaging problems such as imaging denoising, imaging deconvolution, imaging inpainting, etc. Recently, it has been introduced in MR reconstruction successfully. The CP algorithm solves an optimization problem simultaneously with its dual, which provides a robust convergence check – the duality gap. If we denote the prior information $\lambda\|\Psi m\|_1$ in Eq. (1) as $G(m)$, then with CP algorithm, the solution of Eq. (1) is

$$\begin{cases} d_{n+1} = \frac{d_n + \sigma(A\bar{m}_n - f)}{1+\sigma} \\ m_{n+1} = prox_\tau[G](m_n - \tau A^* d_{n+1}) \\ \bar{m}_{n+1} = m_{n+1} + \theta(m_{n+1} - m_n) \end{cases} \quad (2)$$

where $\sigma$, $\tau$ and $\theta$ are the algorithm parameters, and $prox$ denotes the proximal operator, which can be obtained by the following minimization:

$$prox_\tau[G](x) = \arg\min_z \left\{ G(z) + \frac{\|z-x\|_2^2}{2\tau} \right\} \quad (3)$$

Since it is not easy to choose optimal parameters and transforms, and the condition that makes (3) to have a closed-form solution is not always satisfied in practice, a learnable operator is used to replace the proximal operator and is learned through powerfuldeep networks. Thus the algorithm, called PDHG-CSnet, can be formed as

$$\begin{cases} d_{n+1} = \frac{d_n + \sigma(A\bar{m}_n - f)}{1+\sigma} \\ m_{n+1} = \Lambda(m_n - \tau A^* d_{n+1}) \\ \bar{m}_{n+1} = m_{n+1} + \theta(m_{n+1} - m_n) \end{cases} \quad (4)$$

The parameters $\sigma$, $\tau$ and $\theta$ and the operator $\Lambda$ are all learned by the network. As the image prior $G$ is contained in operator $\Lambda$, the PDHG-CSnet learns the regularization functions including both the transform and regularier through the network. In our work, networks are consisted of blocks with convolutional neural network (CNN).

### 2.2 CP-net: learning data consistency

If we relax the constraint of data consistency $\|Am - f\|_2^2$ in Eq. (1) as $F(Am, f)$, then the solution becomes

$$\begin{cases} d_{n+1} = prox_\sigma[F^*](d_n + \sigma A\bar{m}_n) \\ m_{n+1} = prox_\tau[G](m_n - \tau A^* d_{n+1}) \\ \bar{m}_{n+1} = m_{n+1} + \theta(m_{n+1} - m_n) \end{cases} \quad (5)$$

$F^*$ is the convex conjugate of the function $F(Am, f)$ which can be computed by the Legendre transform. Followed by the PDHG-CSnet, a learned operator is also used to replace $prox_\sigma[F^*]$, then the entire iterations of CP algorithm can be rewritten as

$$\begin{cases} d_{n+1} = \Gamma(d_n + \sigma A\bar{m}_n, f) \\ m_{n+1} = \Lambda(m_n - \tau A^* d_{n+1}) \\ \bar{m}_{n+1} = m_{n+1} + \theta(m_{n+1} - m_n) \end{cases} . \quad (6)$$

The primal proximal $\Lambda$, dual proximal $\Gamma$, parameters $\sigma$, $\tau$ and $\theta$ are all learned from training data. To improve the capacity of the network, the parameters of the CNN in each iteration are different, which makes the network a cascading network. We termed this network as CP-net.

### 2.3 PD-net: learning variable structure

To better utilize the learning capability of deep networks and further improve the reconstruction quality based on CP-net, we break the explicitly-enforced updating structure such that the combinations of the variables were freely learned by the network. This is inspired by the learned primal dual in CT reconstruction [13]. Instead of the hard acceleration step $m_{n+1} + \theta(m_{n+1} - m_n)$, the network can be designed to freely learn at what point the forward operator should be evaluated [14]. Thus, the algorithm, called PD-net, is formulated as

$$\begin{cases} d_{n+1} = \Gamma(d_n, Am_n, f) \\ m_{n+1} = \Lambda(m_n, A^* d_{n+1}) \end{cases} \quad (7)$$

### 2.4   Network architecture and training

The entire architecture of primal dual networks and one iteration block of PDHG-CSnet, CP-net and PD-net are illustrated in Fig. 1. The primal and dual iterations have the same architecture with three convolutional layers in each block of CP-net and PD-net. To train the network more easily, we made it a residual network. The convolutions are all 3×3 pixels in size, and for CP-net, the number of channels is 2-32-32-2 in each primal update and 4-32-32-2 in each dual update, whereas for PD-net, the number of channels is 4-32-32-2 in each primal update and 6-32-32-2 in each dual update. The output has two channels representing the real and imagery parts of the data as MR data is complex-value. We set the number of iterations to be 10 in all three networks, and the non-linear operator is chosen to be Rectified Linear Unites (ReLU).

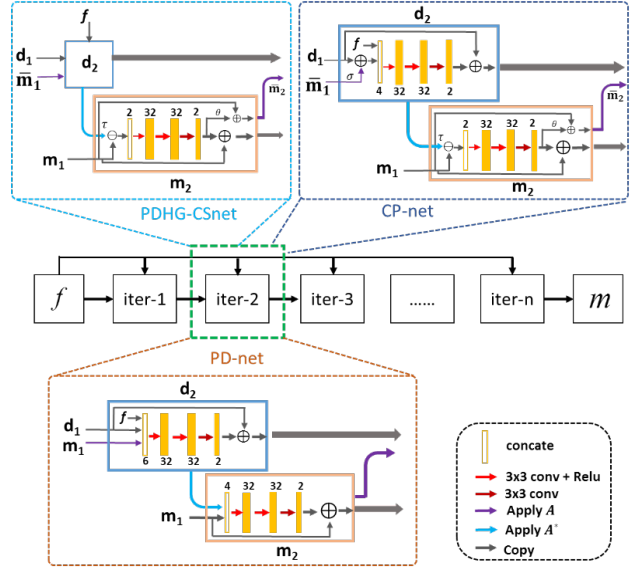

**Fig. 1.** The entire architecture (middle) and one iteration block (top for PDHG-CSnet and CP-net, bottom for PD-net) of the proposed primal dual networks.

In network training, the mean square error (MSE) is chosen as the loss function. Given pairs of training data, the loss between the network output and ground truth is defined as

$$L(\Theta) = \frac{1}{N}\sum_{i=1}^{N}\left\|\widehat{m}(\Theta, f) - m^{ref}\right\|_2^2 \quad (8)$$

where $\hat{m}(\theta, f)$ is the network output based on network parameter $\theta$ and undersampled k-space data $f$, $m^{ref}$ is the corresponding ground truth.

We trained the networks by minimizing the loss function defined above using the ADAM optimizer in TensorFlow. And the trainings were performed on an Ubuntu 16.04 LTS (64-bit) operating system equipped with a Tesla TITAN Xp Graphics Processing Unit (GPU, 12GB memory) with CUDA and CUDNN support.

### 2.5 Relations between networks

PDHG-CSnet and CP-net can be considered as conventional model-driven deep learning methods which unroll the CP algorithm to the deep network. Like previous model-driven methods such as ADMM-net [10] and variational network [9], PDHG-CSnet learns the regularization term (regularization parameters, transformation and regularizer) through networks. However the major difference among these model-driven methods lies in the architectures derived from different optimization algorithms. Whereas CP-net learns the data consistency as well, which relaxes the constraint of data fidelity and makes the reconstruction model more general.

PD-net further relaxes the variable structure constraints based on CP-net, which makes it neither a typical model-driven method nor a purely data-driven method. PD-net learns the reconstruction model with CNN unites updating in k-space and image domain alternatively, which is similar to data-driven method with cross domain learning such as KIKI-net [5]. However, PD-net is derived from a primal dual algorithm with mathematically convergence guarantee, and the formulation is obtained by relaxing the constraints in a specific reconstruction model (1), which are the characteristics of model-driven methods.

## 3 Experiments

We trained the networks using in-vivo MR datasets. Overall 200 fully sampled multi-contrast data from 2 subjects with a 3T scanner (MAGNETOM Trio, SIEMENS AG, Erlgen, Germany) were collected and informed consent was obtained from the imaging object in compliance with the IRB policy. The fully sampled data was acquired by a 12-channel head coil with matrix size of 256×256 and combined to single-channel data and then retrospectively undersampled using Poisson disk sampling mask. After normalization and image augmentation, we got 1600 k-space datasets, where 1400 for training and 200 for validation. We have tested the proposed methods on 7 human brain datasets acquired from three different commercial 3T scanners (SIEMENS AG, Erlgen, Germany; GE Healthcare, Waukesha, WI; United Imaging Healthcare, Shanghai, China).

As the constraints in the specific model (1) are gradually relaxed, the quality of the reconstruction gets better, which is shown in Fig. 2. From PDHG-CSnet to PD-net, the reconstruction model becomes more general and the image quality gradually improves. Nevertheless, the required training set is expected to increase to achieve the optimal performance, which is shown in Fig. 3. The CP-net produces image quality comparable to PD-net with a small training size. On the other hand, as the training size increases, PD-net achieves better quality than others. The performance of a method with more relaxed constraints changes more with increased training data, which is indicated by the variations of quantitative metrics.

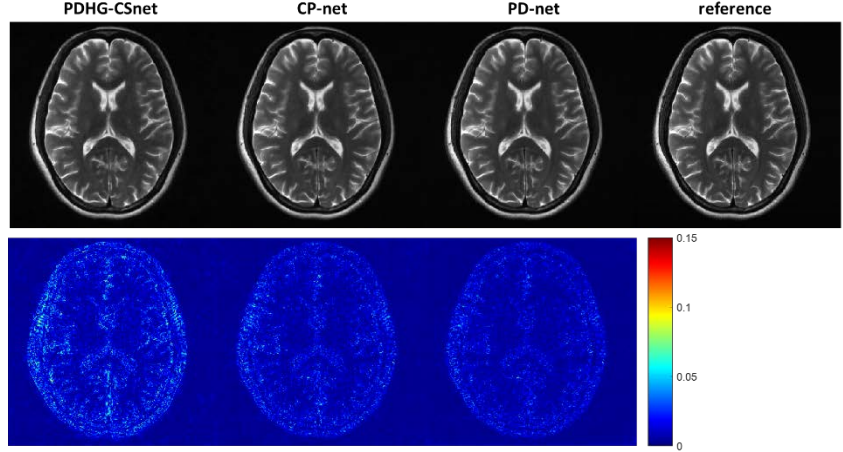

**Fig. 2.** Reconstruction results and the corresponding error maps when gradually relaxing constraints. A 6X Poisson disk sampling mask was used on an axial data from the UIH scanner.

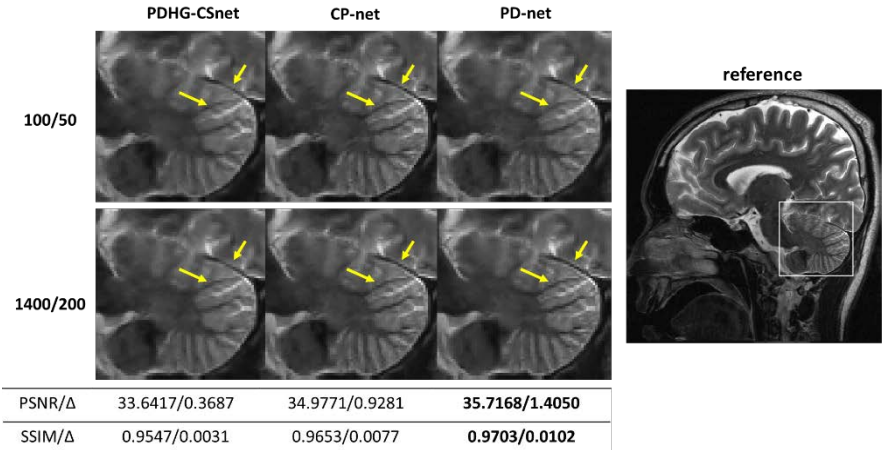

**Fig.3.** The reconstructed zoom-in images of the enclosed part with 6X Poisson disk sampling on a sagittal data from the Siemens scanner. The results with a small training size (100 for training and 50 for validation) are located in the first row and the second row shows the results with more training data (1400 for training and 200 for validation). The quantitative metrics of the results with more training data and the variations on the two datasets are also provided.

We also compared the proposed networks with other reconstruction methods: 1) Rec_PF [15], traditional CSMR reconstruction method to solve problem (1); 2) generic-ADMM-CSnet (ADMM-net) [10], a model-driven deep learning method with the same objective function as PDHG-CSnet; 3) D5C5 [6], a data-driven deep learning method with data consistency; 4) zero-filling, the inverse Fourier transform of undersampled k-space data. The visual comparisons are shown in Fig. 4. The zoom-in images of the enclosed part and the corresponding error maps as well as the quantitative metrics are also provided. Compared to the ADMM-net, CP-net is able to recover more fine details due to the learned data consistency. PD-net achieves better performance than D5C5, although they all update in k-space and image domain alternatively. This is because in D5C5, the data consistency is addressed by $L_2$ Euclidean distance between the estimated and acquired data in k-space, whereas in PD-net, the similarity to the original data is learned by the network, which may be superior to $L_2$ norm. Another possible reason could be that the training data may be not enough for D5C5.

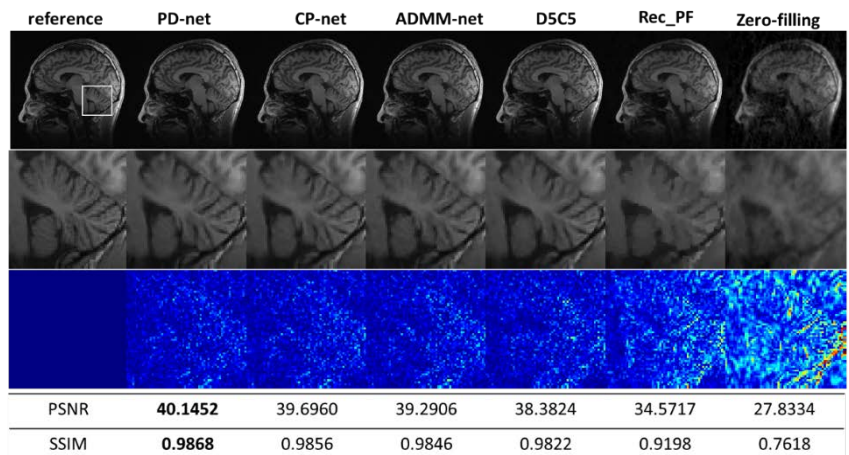

**Fig. 4.** Reconstruction results with 6X Poisson disk sampling on a sagittal data from the GE scanner. The zoom-in images of the enclosed part and the corresponding error maps are provided on the second and third row.

## 4  Conclusion

We developed effective deep networks which integrate classical optimization method and deep network to learn the regularization functions and data consistency at the same time for MR reconstruction. The experimental results on in vivo data demonstrate the effectiveness of the proposed methods in artifacts removal and detail preservation. This work serves as a preliminary attempt to bridge the gap between the model-driven deep learning methods and data-driven deep learning methods. More techniques and properties of the unification of model-driven and data-driven methods should be investigated in the future.

# Acknowledgement

This work was supported in part by the National Science Foundation of China (U1805261, 81729003 and 61871373), the Basic Research Program of Shenzhen (JCYJ20150831154213680), Natural Science Foundation of Guangdong Province (2018A0303130132) and Strategic Priority Research Program of Chinese Academy of Sciences (XDB25000000).